\documentclass[aps,prb,twocolumn,showpacs]{revtex4}

\usepackage{graphicx}

\begin{document}

\title{Molecular adsorption in graphene with divacancy defects}
\author{Biplab Sanyal}
\affiliation{Department of Physics and Materials Science, Uppsala University, Box 530,
  751\,21 Uppsala, Sweden}
\author{Ulf Jansson}
\affiliation{Department of Materials Chemistry, Uppsala University, Box 538,
  751\,21 Uppsala, Sweden}
\author{Olle Eriksson}
\affiliation{Department of Physics and Materials Science, Uppsala University, Box 530,
  751\,21 Uppsala, Sweden}
\author{Helena Grennberg}
\affiliation{Department of Organic Chemistry, Uppsala University, Box 576, SE-751 23 Uppsala, Sweden}

\date{\today}

\begin{abstract}
We have investigated theoretically the adsorption of molecules onto graphene with divacancy defects. Using ab-initio density functional calculations, we have found that O$_{2}$, CO, N$_{2}$, B$_2$ and H$_{2}$O molecules all interact strongly with a divacancy in a graphene layer. Along with a complex geometry of the molecule-graphene bonding, metallic behavior of the graphene layer in presence of CO and N$_{2}$ molecules have been found with a large density of states in the vicinity of the Fermi level suggesting an increase in the conductivity. The adsorption of N$_{2}$ is particularly interesting since the N atoms dissociate in the vicinity of the defects, and take the place where the missing C atoms of the divacancy used to sit. In this way, the defected graphene structure is healed geometrically, and at the same time doped with electron states.
\end{abstract}


%\pacs{75.50.-y,75.30.Cr,71.20.Eh,75.50.Bb}
%75.50.-y Studies of specific magnetic materials
%75.30.Cr Saturation moments and magnetic susceptibilities
%71.20.Eh Rare earth metals and alloys
%75.50.Bb Fe and its alloys
%75.50.Cc Other ferromagnetic metals and alloys

\maketitle
\section{Introduction}
In the last few years, graphene has been considered as one the most sensational materials for its many exotic properties.\cite{geim} Graphene is known theoretically for several decades but the real breakthrough came when it was discovered experimentally in 2004. \cite{novoselov} The electronic properties of this $sp^{2}$-bonded two-dimensional allotrope of carbon have striking similarities with those of massless relativistic Dirac fermions. The linear band dispersion at the so called Dirac point is a special feature of the graphene band structure. Apart from this, graphene has been considered as an excellent candidate for future electronics due to the extraordinarily high mobility of its charge carriers. As today's Si based electronics faces the fundamental limitation of spatial dimension lower than a few nanometers, graphene may stand as a potential alternative.

The electronic structure of graphene may be engineered by the interaction with supporting substrates, gate voltage \cite{pisana} or by chemical doping with molecules. Research on graphene is being pursued in different directions to achieve novel properties. One of the routes is to tune the properties of graphene by controlled defect formation through chemical reactions. In a recent paper \cite{coleman}, it has been shown that chemically treated graphene nanosheets have oxygen-containing defects, which were identified by X-ray absorption spectroscopy (XAS). The oxygen containing groups were introduced  by the acid treatment of the nanosheets, and  follow a reaction pathway similar to what happens for the case of carbon nanotubes with the addition of chemical reagents to carbon network defects in the graphene sheets. Electronic structure calculations revealed that the defect states arise from a network of vacancy defects, where carbon atoms have been removed. %With the aid of electronic structure calculations, the defect states in the XAS spectra were identified as in-plane dangling bond states of $\sigma$ character for a divacancy in the carbon lattice. 
Moreover, it was shown theoretically that the carbon atoms around such a lattice defect show metallic behavior due to the presence of high density of states at Fermi level in contrast to a vanishing density of states (DOS) for a pure graphene sheet. Such a defect would also be very reactive due  to the highly strained structure, and hence efficiently undergo acid-catalyzed water addition, the initial product of which rapidly rearranges to form a pyrane-like structure that possess the characteristics found in the XAS of acid-treated carbon nanosheets (CNS). 

One effective route to induce charge carriers in graphene would be by molecular doping, e.g., by NH$_{3}$, NO$_{2}$ etc. It has been shown by Hall measurements that NO$_{2}$ introduces holes whereas electrons are introduced by doping with NH$_{3}$. This way has opened up an excellent playground for both experimentalists as well as theoreticians. Wehling et al. \cite{wehling} have demonstrated by density functional calculations that a NO$_{2}$ molecule acts as a strong acceptor and produces special features in the density of states (DOS) that is suitable for chemical sensor applications. It was also shown recently \cite{danil} by density functional calculations that chemical functionalization of a bilayer graphene can open up a gap up to 3 eV. It is noteworthy to mention that the detection of gas molecules adsorbed in graphene has been possible recently in experiments by Schedin {\it et al.} \cite{schedin}. These graphene based gas sensors have an extraordinary sensitivity to detect the change in local charge carrier concentration by minute amount due to the extremely low noise in graphene. Besides the characterization and modification of the properties of infinite graphene sheets, a considerable effort has been made recently to predict properties of graphene nanoribbons (GNRs). Modification of electronic properties using edge functionalization, substitutional doping and also the transport and magnetic properties of GNRs have been studied in great detail. \cite{gnr} The edges of GNRs resemble the vacancy defects in graphene and hence are relevant for the present discussion.

The experimental information on the acid-catalyzed addition of water to edges and strained defects without dangling bonds, such as the thermally mobile 5/7 membered rings are available from the research on carbon nanotubes (CNTs) and also some transmission electron microscope (TEM) work on graphites. The addition of water is reversible and heating at low pressure removes quite a lot of water molecules to form a very reactive surface, e.g., defects with dangling bonds  that would relax to still quite reactive but more stable 5-8-5 or related lattice defects. The defects in carbon-bond-network has been studied in detail theoretically by Dias. \cite{dias}
%This has been observed by heat-treating acid-treated  "graphene" (CNS) at very low  pressures followed by feeding the dangling bonds a suitable reagent. 
The purpose of this paper is to study the chemical interaction between divacancy defect centers in graphene and chemisorbed molecules with the aid of theoretical analysis. Using density functional based electronic structure calculations, we have analyzed the reaction between water molecules and defect centers and consequently the passivation of these highly reactive centers by several molecular species. 

\section{Computational details}

Here, we have performed {\it ab-initio} electronic structure calculations 
based on the density functional theory. \cite{DFT} The 
Vienna Ab-initio Simulation Package (VASP) \cite{vasp} employing a plane wave basis  was 
used within the all electron projector augmented wave
(PAW) method. The generalized gradient approximation (GGA) was used as the exchange-correlation functional. GGA reproduces very well the C-C bond length (1.42 \AA ) in graphene. The kinetic energy cut-off was chosen to be 400\,eV. We have optimized the geometries by reducing the Hellman-Feynman forces down to 0.01 eV/ \AA.

We have considered five different molecules and studied their interactions with a divacancy center (e.g. as shown by the C network in Fig.2a) in a graphene sheet. The molecular species we have considered are O$_{2}$, N$_{2}$, B$_{2}$, CO and H$_{2}$O molecules. The graphene sheet was modeled with a 8x8 supercell in the lateral direction with a 15 {\AA} vacuum in the vertical direction. This is a relatively large lateral supercell to reduce the interaction of the adsorbed molecules with their periodic images and can be approximated well by isolated ones.

From the total energy calculations, we have extracted the chemisorption energies for all the molecules considered. The chemisorption energy of the molecule is calculated as 
\begin{equation}
E^{ch}=E_{DV+M}-E_{DV}-E_{M},
\end{equation}
where $E_{DV+M}$, $E_{DV}$ and $E_{M}$ denote the total energies of a graphene sheet with a divacancy and the molecule, a graphene sheet with a divacancy, and a free molecule, respectively. The calculated values of $E^{ch}$ are -8.44 eV, -4.53 eV, -13.83 eV and -3.86 eV respectively for O$_{2}$, N$_{2}$, B$_{2}$ and CO molecules. All these numbers are quite large and they indicate a very strong binding between the molecules and graphene divacancies. A large binding energy between O$_{2}$ molecule and defected graphene indicates it is rather easy to oxidize defected graphene. In case of B$_{2}$, a very large value may arise from the fact that this molecule doesn't form naturally as a diatomic molecule.

\section{Results and discussions}
\subsection{Geometry}

\begin{figure}[hbp]
\begin{center}
\includegraphics[scale=0.51]{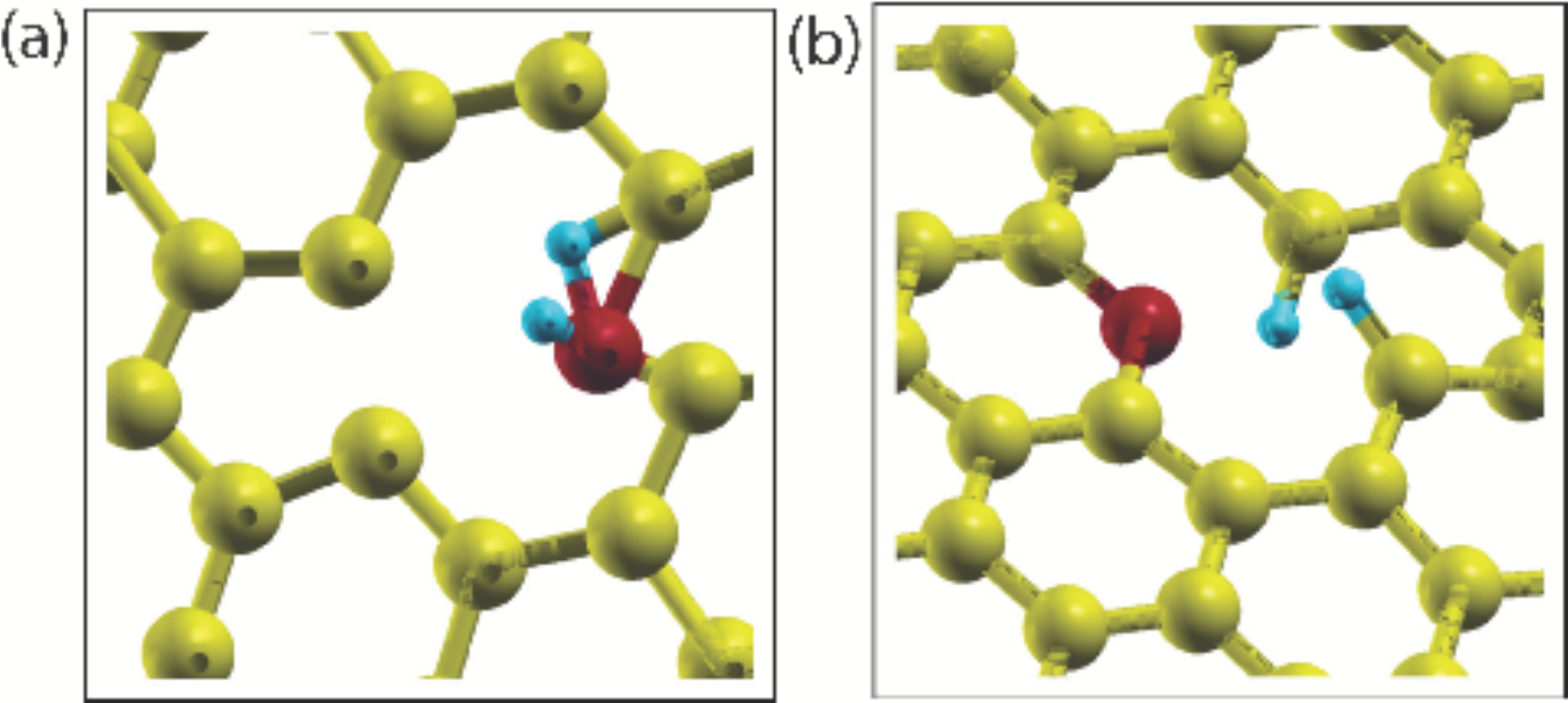}
\caption{(Color online) (a) Initial structure of H$_{2}$O, (b) final structure after geometry optimization. C, O, and H atoms are shown in yellow (light gray), red (dark gray) and turquoise (medium gray) colors respectively.}
\label{fig1}
\end{center}
\end{figure}

Analysis by chemical models suggests that a defect lacking two carbon atoms forming a 5-8-5 lattice defect is more probable than a single vacancy. These defect centers are highly reactive as the carbon atoms at the defect edges have dangling bonds. When this non-perfect induced graphene is treated with water, chemical rearrangement occurs as shown in Fig.~\ref{fig1}. In our study we have used a H$_{2}$O molecule as shown in Fig.~\ref{fig1}(a) in the beginning of the simulation. %In the absence of any H or OH group, the four C atoms around the divacancy hole have dangling bonds. 
In the presence of H$_{2}$O, one C atom at one side of the defect is saturated by H and the other C atom is saturated by OH. The two C atoms at the remaining pentagon of the lattice defect are left untouched. This is one of the probable configurations existing in the reaction. The final configuration after geometry optimization is shown in Fig.~\ref{fig1}(b). The initial quite strained phenolic product is rearranged to a more relaxed pyrane-type structure, with the O atom in a position that produces a hexagonal pattern and the two hydrogen atoms saturating the remaining two carbon atoms of the original defect.
The H atoms are aligned at two opposite faces of the graphene sheet making an angle of around 28$^{\circ}$ to the graphene plane. The final structure is very stable, as all the dangling bonds are saturated.

\begin{figure}[h]
\begin{center}
\includegraphics[scale=0.6]{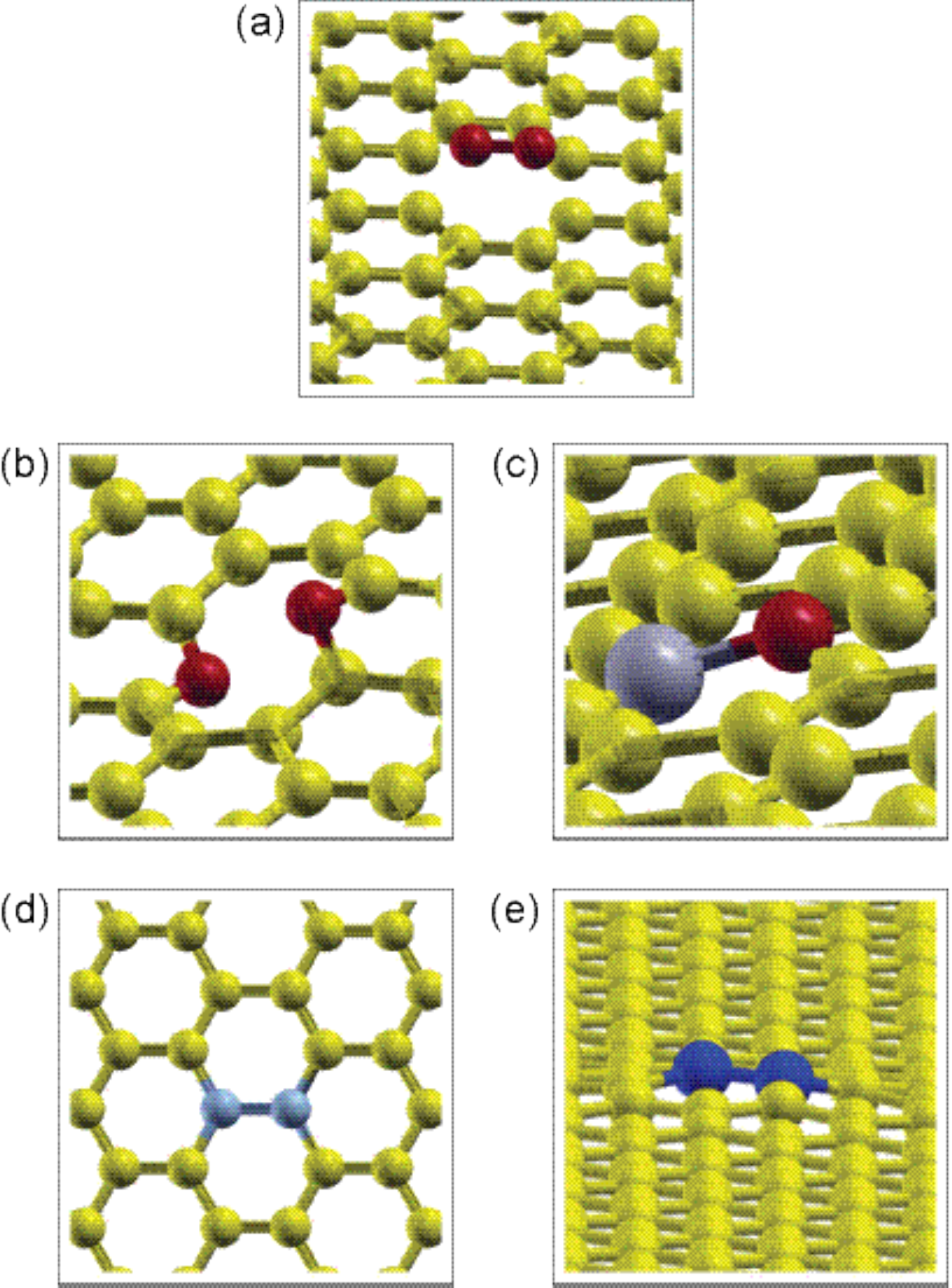}
\caption{(Color online) (a) Initial structure for O$_{2}$, CO, N$_{2}$ and B$_{2}$ molecules. Final relaxed structures for (b) O$_{2}$, (c) CO, (d) N$_{2}$ and (e) B$_{2}$ molecules. In all panels, C atoms are shown in yellow (light shade). In (b), O atoms are shown in red (dark gray), in (c), C and O atoms of CO are in gray (gray) and red (dark gray) respectively, in (d), N atoms are in turquoise (medium gray) and in (e), B atoms are shown in blue (blackish gray).} 
\label{fig2}
\end{center}
\end{figure}

%Now consider that the system undergoes heat treatment. The systems gets rid of O and H atoms and becomes reactive once again in presence of dangling bonds. % 
Similar to the case of H$_{2}$O,  a graphene with unsaturated dangling bonds  will immediately absorb any molecular center close to it. We have considered O$_{2}$, CO, B$_2$ and N$_{2}$ molecules interacting with the lattice defect in the C network of graphene. Fig.~\ref{fig2}(a) shows the initial configuration of the simulation. The molecule is kept at a horizontal position above the vacancy region and the geometry of the system was optimized by minimizing the forces. The final optimized configurations are shown in Fig.~\ref{fig2}(b), \ref{fig2}(c), \ref{fig2}(d) and \ref{fig2}(e) for O$_{2}$, CO, N$_{2}$ and B$_2$ molecules respectively.

The three molecular species behave differently while being adsorbed in the graphene sheet. The O$_{2}$ molecule gets dissociated into two O atoms bound to two pairs of C atoms around the divacancy. The O atoms are moved in opposite directions out of the graphene plane, a situation similar to that of H atoms described in Fig.~\ref{fig1}. For the case of CO, the C and O atoms still remain bonded to each other, while being bonded to two pairs of carbon atoms. Moreover, the CO molecule doesn't remain in the plane of graphene. The C atom of the CO molecule stays in the plane making a usual hexagonal planar structure while the O atom moves a little bit out of the plane. The N$_{2}$ molecule is completely adsorbed in the plane of graphene. The two N atoms are substituted in the normal positions of a graphene network. So, a graphene sheet with substitutional N atoms is achieved in this process. This may serve as a route to dope N atoms in the graphene network and hence introduce extra charge carriers to achieve novel transport properties. In case of B$_{2}$, the B atoms are buckled out of the graphene plane by about 0.55 {\AA}.

\subsection{Electronic structure}

\begin{figure}[h]
\begin{center}
\includegraphics[scale=0.40]{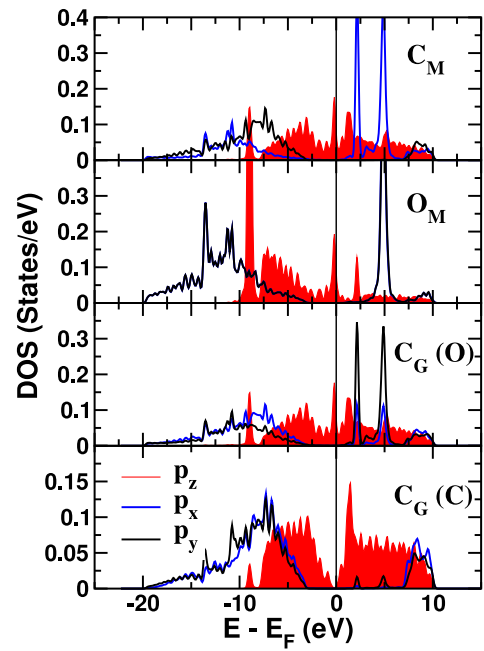}
\caption{(Color online) DOS for CO on graphene. In the panels, C$_{M}$, O$_{M}$, C$_{G}$ (O) and C$_{G}$ (C) indicate DOSs for C atom in CO, O atom in CO, C atom in the graphene sheet connected to O and C atom in the graphene sheet connected to C atom of CO molecule respectively. For all the cases, DOSs projected on different p orbitals are shown. }
 \label{fig3}
\end{center}
\end{figure}

We now present the electronic structure of graphene with CO, N$_{2}$ and B$_{2}$ chemisorbed to the lattice defect.. The case for the CO molecule is shown in Fig.~\ref{fig3}. C$_{M}$, O$_{M}$, C$_{G}$ (O) and C$_{G}$ (C) indicate the C atom in CO, the O atom in CO, the C atom in the graphene sheet connected to O and the C atom in the graphene sheet connected to the C atom of CO molecule, respectively. Projected DOSs on different atoms and orbitals show different behaviors. The p$_{z}$ orbital-projected DOSs of C (C$_{M}$) and O (O$_{M}$) atoms in the CO molecule and the C (C$_{G}$(O)) atom in the graphene sheet connected to the O atom of CO, have a rather sharp peak at the Fermi level (E$_{F}$). This feature is very similar to the mid-gap states observed \cite{wehling, coleman} in graphene with defects. The C (C$_{G}$ (C)) atom in the graphene layer connected to the C atom of CO has a usual graphene-like behavior, i.e., a vanishing DOS at E$_{F}$. For this atom the electronic structure is actually quite similar to that of defect free graphene, with the p$_x$ and p$_y$ orbitals being almost degenerate forming an occupied $\sigma^{}$ state, whereas the p$_z$ orbital forms a $\pi$ and $\pi^*$ complex just below and just above the Fermi level, respectively. This picture is somewhat different for the other DOS curves in Fig.1, and it is clear that for C$_{M}$ and C$_{G}$(O) the p$_x$ and p$_y$ orbitals are no longer degenerate, since symmetry is lowered due to the geometric reconstructions.

\begin{figure}[h]
\begin{center}
\includegraphics[scale=0.36]{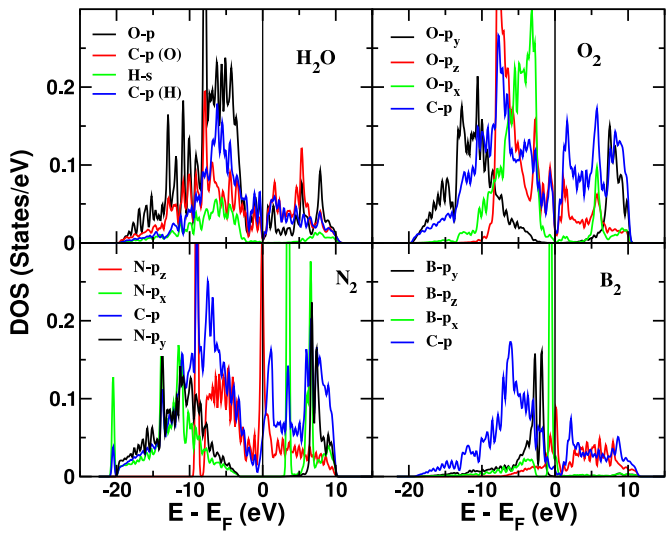}
\caption{(Color online) DOSs for (top left) H$_{2}$O, (top right) O$_{2}$, (bottom left) N$_{2}$ and (bottom right) B$_{2}$ molecules are shown. In the top left panel, C-p(O) and C-p(H) indicate the p orbitals of the C atoms of graphene bonded to O and H respectively. For O$_{2}$, N$_{2}$ and B$_{2}$, C-p indicates the p orbital projected DOS for one of the C atoms connected to the molecule.}
\label{fig4}
\end{center}
\end{figure}

In Fig.~\ref{fig4}, we have shown the atom and orbital projected DOSs for N$_{2}$, B$_2$,  H$_{2}$O and O$_{2}$ molecular species adsorbed on a divacancy in graphene. For N$_{2}$ adsorption, we observe a large DOS very close to E$_{F}$. In this case, E$_{F}$ lies almost at a very sharp peak primarily of N p$_{z}$ character. One  might suspect that this could lead to a Stoner instability and a spin-polarized ground state. In order to investigate this we also performed spin-polarized calculations. However, in these calculations the magnetic moment vanishes to result in a spin-degenerate state. The results of N$_2$ adsorbed are nevertheless very interesting since the N molecules, after geometry optimization are found to take positions to generate a perfect honeycomb lattice. Hence, after N$_2$ adsorption the divacancy defect has healed, generating a perfect graphene geometry, albeit with N atoms replacing some of the C atoms. This naturally provides electron doping into the lattice, and our results may hence provide a recipe for doping graphene with electrons. One may envision creating divacancy (or similar) defects of graphene by ion- or electron-irradiation under vacuum conditions, and afterwards flush the sample with N$_2$ gas, and in this way provide a graphene like material, with electrons doped into the conduction band. In case of B$_2$, a very sharp resonance of p$_{x}$ character is observed just below the Fermi energy followed by a smaller peak of p$_{z}$ character.
The large DOS in the vicinity of Fermi level is absent for the cases of H$_{2}$O and O$_{2}$. In these cases, a pseudogap at the Fermi level is produced similar to the case of pure graphene. The DOS of the H$_{2}$O adsorbed system is particularly interesting having in mind the experimental study of Ref.\onlinecite{coleman}. In the present study, peaks in the C and O atom projected DOSs are observed around 4 eV above the Fermi level. In the paper mentioned above, a peak in the XAS spectra was observed around 3-4 eV above the threshold. The occurrence of this peak was argued to be due to the presence of vacancy defects. Here, a similar observation is made in presence of the pyrane-type saturated lattice defect. One may conclude that unsaturated dangling bond vacancy defects as well as saturated bond vacancies saturated by reaction with water may be the cause for the modification of the electronic structure of graphene sheets.

\section{Conclusions}
 
In conclusion, we have studied here the adsorption of different molecular 
species to a graphene with a divacancy, correpsponding to a highly likely 5-8-5 lattice defect. The calculated chemisorption energies for the molecules are quite large, with a magnitude which is of order 3-13 eV. Our ab-initio density functional calculations indicate that N$_{2}$ is adsorbed in the plane of the graphene network as substitutional impurities, whereas H$_{2}$O, CO and O$_{2}$ have complex geometrical constructions at the divacancy site. The B$_{2}$ molecule more or less retains its bond length as in the isolated molecular state and is strongly bonded to the divacancy center in a position a little bit off the graphene plane. The significant observation is that the inclusion of N$_{2}$ in the graphene plane produces a perfect honeycomb lattice with electrons doped into the conduction band. A recipe for doing this experimentally is suggested.  

\section{Acknowledgements}

We gratefully acknowledge financial support from the Swedish Research Council (VR), 
Swedish Foundation for Strategic Research (SSF), G\"oran 
Gustafssson foundation and a KOF grant from Uppsala University. We also acknowledge Swedish National Infrastructure for Computing (SNIC) for the allocation of time in high performance supercomputers.

%\begin{thebibliography}{27}
%\expandafter\ifx\csname natexlab\endcsname\relax\def\natexlab#1{#1}\fi
%\expandafter\ifx\csname bibnamefont\endcsname\relax
 % \def\bibnamefont#1{#1}\fi
%\expandafter\ifx\csname bibfnamefont\endcsname\relax
 % \def\bibfnamefont#1{#1}\fi
%\expandafter\ifx\csname citenamefont\endcsname\relax
 % \def\citenamefont#1{#1}\fi
%\expandafter\ifx\csname url\endcsname\relax
 % \def\url#1{\texttt{#1}}\fi
%\expandafter\ifx\csname urlprefix\endcsname\relax\def\urlprefix{URL }\fi
%\providecommand{\bibinfo}[2]{#2}
%\providecommand{\eprint}[2][]{\url{#2}}

%\end{thebibliography}

\end{document}